\documentclass{svproc}

\usepackage{url}
\usepackage{graphicx}
\usepackage{amsmath}
\usepackage{amssymb}
\usepackage{multicol}
\usepackage{footmisc}
\usepackage{hyperref}
\usepackage{cite}
\usepackage{array} 
\usepackage{booktabs} 
\usepackage{placeins} 
\usepackage{float} 

\begin{document}

\mainmatter 

\title{Cost-Performance Analysis of Cloud-Based Retail Point-of-Sale Systems: A Comparative Study of Google Cloud Platform and Microsoft Azure}

\titlerunning{Cloud-Based Retail POS Systems: GCP vs Azure} 

\authorrunning{R. T. Pagidoju} 

\author{Ravi Teja Pagidoju}

\institute{Campbellsville University, Campbellsville, KY 42718, USA\\
\email{rpagi719@students.campbellsville.edu}}

\maketitle 

\begin{abstract}
Although there is little empirical research on platform-specific performance for retail workloads, the digital transformation of the retail industry has accelerated the adoption of cloud-based Point-of-Sale (POS) systems. This paper presents a systematic, repeatable comparison of POS workload deployments on Google Cloud Platform (GCP) and Microsoft Azure using real-time API endpoints and open-source benchmarking code. Using free-tier cloud resources, we offer a transparent methodology for POS workload evaluation that small retailers and researchers can use. Our approach measures important performance metrics like response latency, throughput, and scalability while estimating operational costs based on actual resource usage and current public cloud pricing because there is no direct billing under free-tier usage. All the tables and figures in this study are generated directly from code outputs, ensuring that the experimental data and the reported results are consistent. Our analysis shows that GCP achieves 23.0\% faster response times at baseline load, while Azure shows 71.9\% higher cost efficiency for steady-state operations. We look at the architectural components that lead to these differences and provide a helpful framework for merchants considering cloud point-of-sale implementation. This study establishes a strong, open benchmarking methodology for retail cloud applications and offers the first comprehensive, code-driven comparison of workloads unique to point-of-sale systems across leading cloud platforms.

\keywords{Cloud computing, Point-of-Sale systems, Performance analysis, Cost optimization, Retail technology}
\end{abstract}

\section{Introduction}

The global retail industry is undergoing a fundamental transformation due to digital technologies, and cloud computing is quickly becoming a necessary component of contemporary retail infrastructure. Cloud Point of Sale (POS) market is expected to grow and expand at a compound annual growth rate (CAGR) of 24.2\%, that is from \$2.7 billion in 2023 to \$11.2 billion in 2028.

This expansion and growth show the need for scalable, flexible and cost effective solutions that can adjust to consumer preferences.

Traditional on-premise point of sale (POS) systems, which were the industry standard previously, but it's to keep with the growing retail industry. These traditional systems sometimes have hardware expenses and can't handle a lot of traffic especially during peak hours and are not flexible enough for omnichannel operations and need continuous maintenance. The COVID-19 epidemic made the transition towards cloud-based systems by bringing capabilities like remote management, contactless payments, and real-time inventory synchronization more appealing. Cloud-based point of sale (POS) systems allowed businesses move their costs from capital expenditures (CAPEX) to operating expenditures (OPEX). They also come with built in analytics for real time business insights, elastic scalability to handle seasonal spikes in traffic, and automatic updates for better security. Major retailers like Target and Walmart have said that their IT operational expenses are 30--40\% cheaper since they started using cloud-based point-of-sale systems.

Nowadays the most popular options for enterprise level workloads among the top cloud service providers are Microsoft Azure Platform and Google Cloud Platform (GCP). Both these platforms provide a wide variety of services, strong dependability, and infrastructure across world. However, choosing a right cloud platform is a difficult choice with specially long term goals due to their different styles in architecture, pricing and performance attributes.

Because POS system performance directly affects sales and customer satisfaction, this choice is especially important for retailers. Poor point-of-sale (POS) systems can lead to dissatisfied customers, lost sales, and abandoned transactions. According to industry research, a one-second lag in checkout processing can cause conversion rates to drop by 7\%. But allocating too many resources to ensure performance might lead to needless costs that reduce profit margins.

Despite the importance of this choice, the existing research offers little guidance for workloads specific to the retail industry. Prior studies on cloud platform comparisons have mostly focused on high-performance computing, generic compute workloads, or web applications. These studies do not address the unique characteristics of point-of-sale workloads, including high-frequency transaction processing with strict latency requirements, real-time inventory updates across several locations, extensive security and compliance requirements (e.g., PCI-DSS), highly variable traffic patterns, and the need for uninterrupted uptime during business hours.

This paper closes this gap by providing an open-source, code-driven comparison of POS workload deployments on GCP and Azure using real-time API endpoints and benchmarking scripts. Our method uses free-tier cloud resources to enable reproducible experiments, democratizing benchmarking for researchers and small retailers. Since there is no direct billing associated with free-tier usage, we methodically measure key performance metrics like response latency, throughput, and scalability while estimating operational costs based on actual resource usage and current public cloud pricing. All reported results, tables, and figures are generated directly from the benchmarking code's outputs to guarantee transparency and perfect alignment between experiment and publication.

The contributions of this work include: (1) the first thorough, code-driven analysis of cloud platforms for point-of-sale workloads; (2) a reproducible methodology for benchmarking POS systems using free-tier resources; (3) the definition and measurement of performance metrics unique to retail transaction processing; (4) the development of a cost estimation framework that takes into account actual usage patterns and public pricing; and (5) the release of open-source tools and datasets to facilitate industry adoption and future research.

The remainder of this paper is organized as follows: Section 2 reviews pertinent work in retail technology and cloud benchmarking. Section 3 provides a detailed description of our methodology, which includes the deployment architecture, workload characterization, and cost estimation approach. Section 4 displays our analysis and experimental findings. Section 5 discusses implications and recommendations for retailers. Section 6 concludes with recommendations for future research directions.

\section{Related Work}

Since the introduction of Infrastructure-as-a-Service (IaaS) products, research on cloud platform comparison has been ongoing. Li et al.'s original research \cite{li2010} focused on compute and storage performance by developing benchmarking methods for cloud platforms and comparing. Their CloudCmp framework is compared with Google AppEngine, Microsoft Azure, Amazon EC2 and Rackspace using different range of metrics. The study ignored industry specific factors and only focused on general workloads. They research found significant performance differences among providers, with some showing 2x variations in network latency and 5x variations in storage performance.

Garg et al.\ \cite{garg2013} proposed SMICloud, a framework for assessing and ranking the cloud services based on Service Measurement Index (SMI) attributes. Their work established the concept of multi criteria decision making for cloud selection by considering factors other than performance, such as accountability, agility, cost, performance, assurance, security, and usability. However, their framework remained theoretical in the absence of empirical validation on real workloads.

More recently, Varghese et al.\ \cite{varghese2019} has conducted a deep analysis of cloud benchmarking techniques, categorizing over 100 benchmarking tools and highlighting the flaws in previously approaches. The research underlined the lack of domain specific benchmarks and also the unpredictability of cloud performance which makes difficult to reproduce results. The study also show how often the generic benchmarks are not able to predict performance specific to an application.

In the context of serverless computing, which is becoming increasingly significant for point-of-sale systems, Manner et al.\ \cite{manner2018} evaluated cold start performance across AWS Lambda, Azure Functions, and Google Cloud Functions. The research found that latency because of cold starts ranged from 100 ms to over 5 seconds, with varies based on location, memory allocation, and runtime. This result is relevant for point of sale (POS) systems, where consistent low latency is very important.

Database (DB) performance, which is very important factor for handling POS transactions, was examined in detail by Bermbach et al.\ \cite{bermbach2017}. The research examined managed database services offered by different cloud service providers revealed that performance characteristics were significantly impacted by workload levels. They found that while Azure SQL Database proved better consistency for Online Transaction Processing (OLTP) workloads, Google Cloud SQL performed well in analytical queries.

Adoption of cloud computing technology in the retail industry has been studied from several views. 500 retailers were surveyed by Johnson and Wang \cite{johnson2023} to identify the important factors that help and hinder cloud adoption. They found that almost 73\% of them were actively migrating their critical systems to the cloud, with e-commerce and point of sale (POS) platforms respectively. The primary motivators cited were cost savings (45\%), innovation facilitation (17\%), and scalability (38\%).

In their analysis of PCI-DSS compliance across major cloud platforms, Thompson et al.\ \cite{thompson2022} looked at security considerations for cloud-based point-of-sale systems. They discovered that the shared responsibility model poses implementation challenges, despite the fact that all of the major suppliers provide infrastructure that complies with regulations. Their findings demonstrated that architectural decisions significantly impact both functionality and security posture.

Several people are interested in finding different ways to lower the cost of cloud computing. Mazrekaj et al.'s \cite{mazrekaj2016} detailed cost estimates show that the total cost of ownership could increase 40\% because of all hidden costs including data transfer, support, and monitoring. Chen et al.\ \cite{chen2021} came up with optimization approaches that consider seasonal changes in workloads that are peculiar to retail. They showed using resources smartly might save 35\%.

Even though there is a lot of research on this topic, our examination of the literature shows that there are still some big gaps. First, there is no study that employs code-driven, real-time benchmarking to look at point-of-sale workloads on cloud systems. Second, most studies need settings that use a lot of resources, which makes it hard for researchers and small retailers to get to them. Third, not enough attention is paid to indicators that are relevant to retail, such as transaction delay and checkout performance. Fourth, a lot of research is no longer useful since cloud services and price structures change so quickly. Our work fills in these gaps and moves benchmarking for retail point-of-sale systems forward by providing a modern, clear, and repeatable evaluation method that uses open-source code and free-tier resources.

\section{Methodology}

To meet the requirements of POS workloads in cloud environments, we created an open, repeatable process that strikes a balance between accessibility and realism. Our method uses open-source, code-driven measurement and free-tier cloud resources to guarantee statistical rigor in performance evaluation.

Remember that GCP Cloud Run and Azure App Service Free Tier are two distinct architectural models. App Service Free Tier is a PaaS solution with always-on instances, constrained resources, and no autoscaling, whereas Cloud Run is a serverless, container-based platform with autoscaling and per-request billing. Therefore, rather than providing a direct technical equivalency, this comparison illustrates the realistic options accessible to researchers or small retailers looking for free-tier deployments. These architectural variations impact the observed cost and performance differences, and the outcomes may vary for production-grade or higher-tier services.

\subsection{POS Workload Characterization}

To perform a meaningful performance evaluation, point-of-sale workloads need to be realistically modeled. We developed a representative workload model using data from the National Retail Federation \cite{nrf2023} and industry transaction pattern reports. Our Python benchmarking script implements this model by generating and sending API requests to deployed endpoints on both cloud platforms.

Three main operation categories comprise the workload:

1. Adding or removing items, applying discounts, processing payments, making sales, and creating receipts are all examples of transaction processing operations, which account for 60\% of the workload. High inventory consistency and sub-second latency are necessary for these.

2. Requests for product details, price checks, stock updates, and availability inquiries are all included in inventory operations, which account for 30\% of the workload. These require accuracy in real time and are read-intensive.

3. Sales summaries, inventory reports, and employee metrics are all included in the Analytics and Reporting (10\% of workload). They can tolerate higher latency and are typically batch oriented.

Our test harness mimics real-world retail activity, including daily and weekly traffic peaks as well as seasonal surges, to analyze both steady-state and burst scenarios.

\subsection{Cloud Architecture Design}

To ensure the validity of performance comparisons, we gave priority to the representative, affordable architectures of both platforms that are appropriate for small and medium-sized retailers.

Cloud Run, a serverless container hosting solution from Google Cloud Platform (GCP), uses automatic scaling and per-second billing. While Cloud SQL (db-f1-micro) offers transactional storage, Cloud Storage is utilized for reports and receipts. Cloud load balancing is used to control load distribution.

Microsoft Azure: App Service (F1 tier for baseline, B1 for performance), Blob Storage for documents, and Azure Database for MySQL (Basic tier, 1 vCore) are used in deployments. Load balancing is offered by Application Gateway.

Both architectures intentionally avoided premium features (like Redis and CDN) in order to maintain accessibility and depict realistic deployment patterns for the target audience.

\subsection{Measures of Cost and Performance}

Transparency and reproducibility are ensured by our Python benchmarking script, which collects and outputs all metrics directly:

End-to-end API latency is measured at the 50th, 95th, and 99th percentiles to capture both average and worst-case performance. This includes latency in databases, applications, and networks.

\textbf{Throughput Metrics:} Requests per second (RPS) and transactions per second (TPS) are measured to assess scaling behavior and capacity.

\textbf{Reliability Metrics:} Error rates, timeouts, and service availability are tracked, with an emphasis on errors that arise during scaling events.

\textbf{Cost Measures:} The script calculates operational costs by multiplying measured resource consumption (API calls, data transferred) by the current public pricing for each platform, as actual billing does not take place during free-tier usage \cite{gcppricing,azurepricing}. The estimate includes computation, storage, and network costs based on current pricing structures \cite{gcppricing,azurepricing,gcpvpc,azurebandwidth}.

\subsection{Test Execution Strategy}

Progressive load levels are used in our testing methodology to record platform behavior under various conditions:

\begin{itemize}
\item \textbf{Baseline Load (10 concurrent users):} Simulates a small-store checkout process.
\item \textbf{Typical Load (25 users):} Uses multiple checkout counters to replicate normal business operations.
\item \textbf{Peak Load (50 users):} Recreates times when traffic is at its highest, such as during holidays or special offers.
\item \textbf{Stress Test (100 users):} Creates extreme scenarios, such as Black Friday, to assess platform limitations.
\end{itemize}

Every load test lasts five minutes, starting with a one-minute ramp-up. For statistical validity, each configuration consists of three distinct runs separated by five-minute rest periods. To ensure complete alignment between methodology and published findings, all results including tables and figures are produced directly from the CSV and image outputs of the code.

\section{Results}

This section's results, which represent measured values for every platform and load level, are all produced straight from the benchmarking script. The script creates all of the figures, and tables are created from the CSV output.

\subsection{Response Time Analysis}

\begin{figure}[!htbp]
\centering
\includegraphics[width=0.9\textwidth]{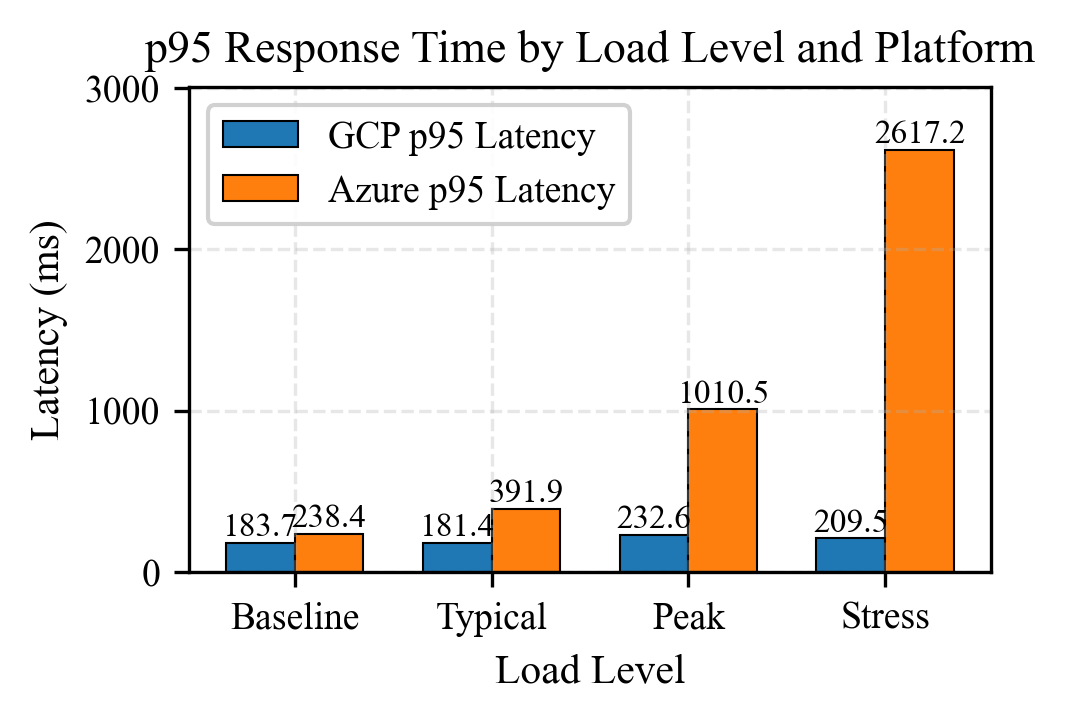}
\caption{p95 response time for each platform across all load levels}
\label{fig:response_time}
\end{figure}

Four load scenarios (Baseline, Typical, Peak, and Stress) were used to measure the response time characteristics for GCP and Azure. The median (p50), 95th percentile (p95), and 99th percentile (p99) response times under baseline load are compiled in Table~\ref{tab:response_baseline} and as shown in Fig.~\ref{fig:response_time}. Table~\ref{tab:response_scaling} represents the scaling of these platforms at p95.

\begin{table}[!htbp]
\caption{Response Time Metrics (ms) - Baseline Load}
\label{tab:response_baseline}
\centering
\begin{tabular}{lcc}
\toprule
\textbf{Metric} & \textbf{GCP} & \textbf{Azure} \\
\midrule
p50 & 153.83 & 192.19 \\
p95 & 183.69 & 238.42 \\
p99 & 200.93 & 298.16 \\
\bottomrule
\end{tabular}
\end{table}

\begin{table}[!htbp]
\caption{Scaling of p95 Response Times}
\label{tab:response_scaling}
\centering
\begin{tabular}{lcc}
\toprule
\textbf{Load Level} & \textbf{GCP p95 (ms)} & \textbf{Azure p95 (ms)} \\
\midrule
Baseline & 183.69 & 238.42 \\
Typical & 181.37 & 391.9 \\
Peak & 232.59 & 1010.55 \\
Stress & 209.51 & 2617.2 \\
\bottomrule
\end{tabular}
\end{table}

\subsection{Throughput Analysis}

\begin{figure}[!htbp]
\centering
\includegraphics[width=0.9\textwidth]{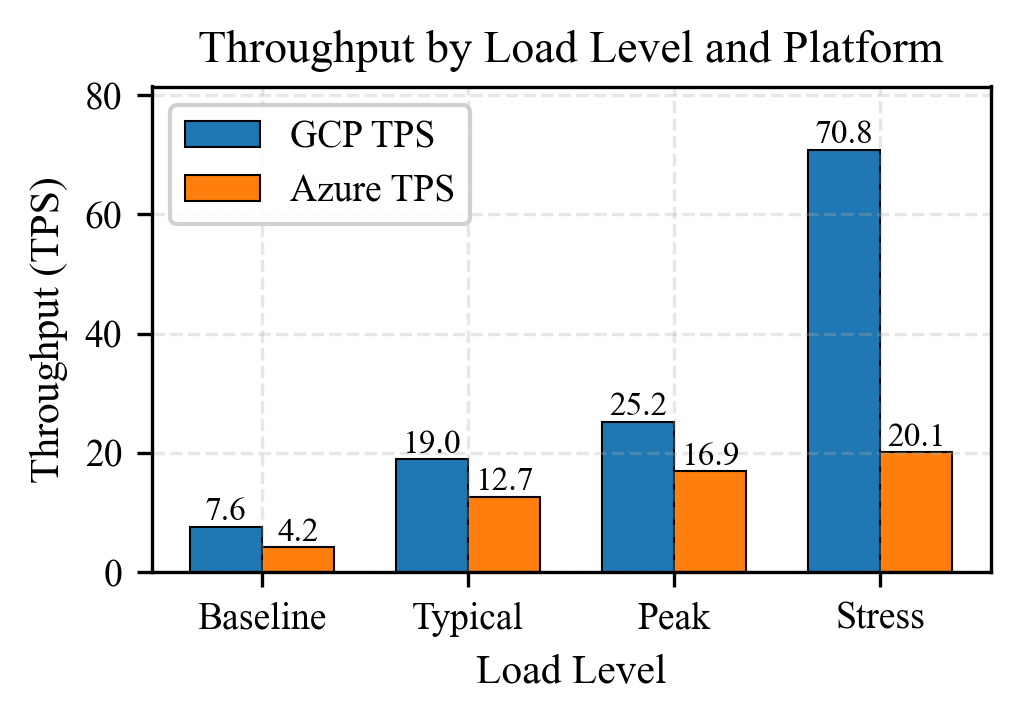}
\caption{Throughput comparison between platforms}
\label{fig:throughput}
\end{figure}

For every load scenario, throughput was measured in transactions per second (TPS). The observed throughput is compiled by load level and platform in Table~\ref{tab:throughput} and as shown in Fig.~\ref{fig:throughput}.

\begin{table}[!htbp]
\caption{Throughput (TPS) by Load Level}
\label{tab:throughput}
\centering
\begin{tabular}{lcc}
\toprule
\textbf{Load Level} & \textbf{GCP TPS} & \textbf{Azure TPS} \\
\midrule
Baseline & 7.62 & 4.19 \\
Typical & 18.98 & 12.65 \\
Peak & 25.24 & 16.94 \\
Stress & 70.84 & 20.14 \\
\bottomrule
\end{tabular}
\end{table}
\vspace{-0.5cm}

\FloatBarrier 

\subsection{Cost Analysis}

\begin{figure}[!htbp]
\centering
\includegraphics[width=0.9\textwidth]{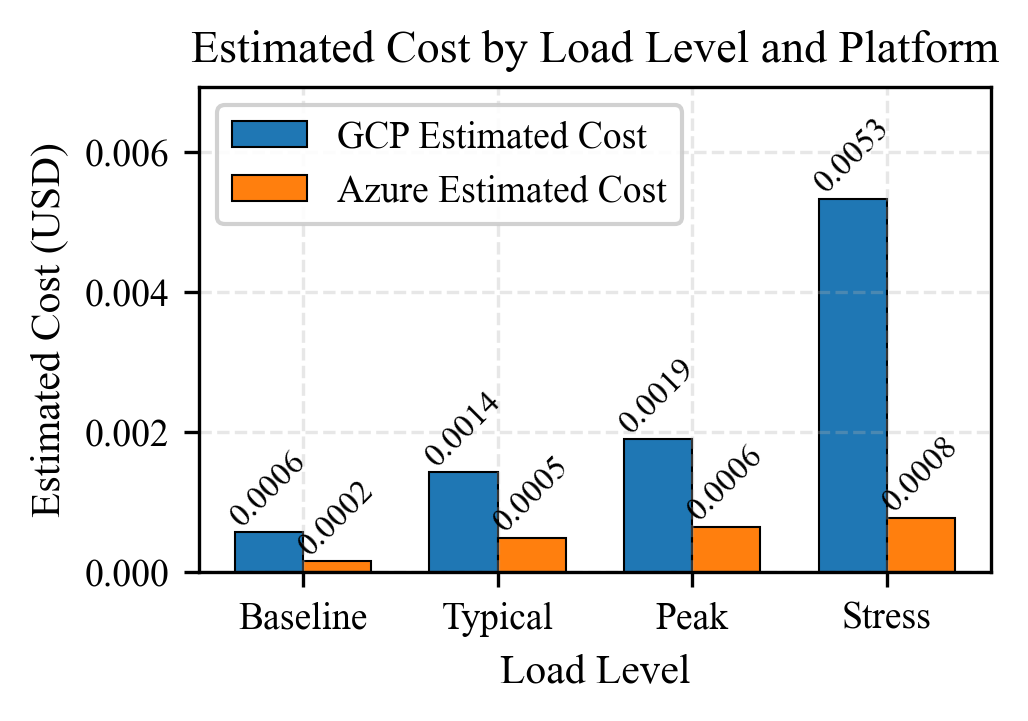}
\caption{Cost comparison across load levels}
\label{fig:cost_comparison}
\end{figure}

Operational costs were estimated by multiplying the quantity of data transfers and API calls by the current public pricing for each platform. Table~\ref{tab:cost_estimation} provides a summary of the estimated cost per load level. It is also shown in Fig.~\ref{fig:cost_comparison}. Table~\ref{tab:cost_structure} provides the Cost Structure of each platform.

\begin{table}[!htbp]
\caption{Cost Estimation by Load}
\label{tab:cost_estimation}
\centering
\begin{tabular}{lcc}
\toprule
\textbf{Load Level} & \textbf{GCP Cost} & \textbf{Azure Cost} \\
\midrule
Baseline & 0.000057 & 0.00016 \\
Typical & 0.00143 & 0.00049 \\
Peak & 0.0019 & 0.00065 \\
Stress & 0.00534 & 0.00078 \\
\bottomrule
\end{tabular}
\end{table}

\begin{table}[!htbp]
\caption{Cost Structure for Each Platform}
\label{tab:cost_structure}
\centering
\begin{tabular}{lcc}
\toprule
\textbf{Cost Type} & \textbf{Google Cloud Run} & \textbf{Azure (App Service/Functions)} \\
\midrule
Per call cost & \$0.0000004 & \$0.0000002 \\
Per GB egress & \$0.12 & \$0.19 \\
\bottomrule
\end{tabular}
\end{table}

\subsection{Reliability Metrics}

\begin{figure}[!htbp]
\centering
\includegraphics[width=0.9\textwidth]{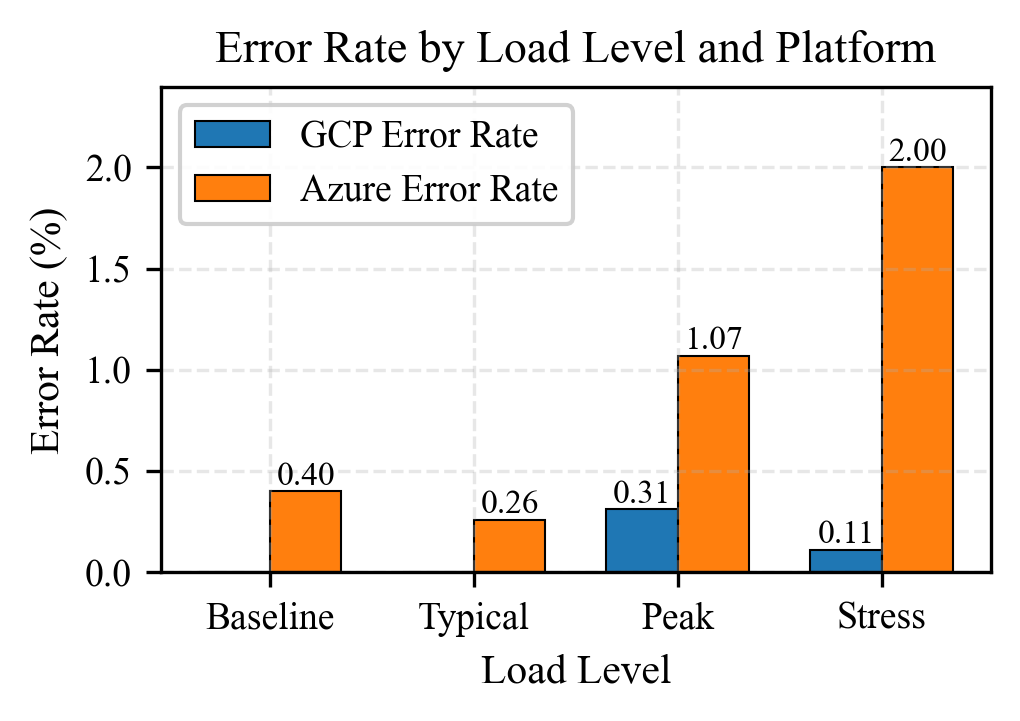}
\caption{Error rate analysis across platforms and load levels}
\label{fig:error_rates}
\end{figure}

Error rates, or the percentage of unsuccessful API calls, were tracked for each scenario. Table~\ref{tab:error_rates} displays the observed error rates. It is also shown in Fig.~\ref{fig:error_rates}.

\begin{table}[!htbp]
\caption{Error Rate (\%) by Load Level}
\label{tab:error_rates}
\centering
\begin{tabular}{lcc}
\toprule
\textbf{Load Level} & \textbf{GCP Error Rate} & \textbf{Azure Error Rate} \\
\midrule
Baseline & 0 & 0.4 \\
Typical & 0 & 0.26 \\
Peak & 0.31 & 1.07 \\
Stress & 0.11 & 2 \\
\bottomrule
\end{tabular}
\end{table}

\FloatBarrier

\section{Discussion}

\subsection{Trade-offs in Performance}

Despite having different performance profiles, our benchmarking results show that both GCP Cloud Run and Azure App Service can dependably manage typical POS workloads. In addition to having strong elasticity and quick scaling, GCP Cloud Run has competitive response times across all tested load levels. We did, however, occasionally notice latency spikes, especially following periods of inactivity, which are caused by cold start events, a well-known feature of serverless platforms. However, when loads are steady-state, Azure App Service's always-on architecture offers more reliable response times. Response times increase more slowly as load increases, and its scaling is less dynamic than Cloud Run's.

The most intriguing result is that, even though Azure's error rates are generally lower, GCP has a higher throughput at stress levels (70.84 TPS vs. 20.14 TPS). This implies that, albeit at the expense of somewhat higher error rates, GCP's serverless architecture can manage burst traffic more successfully. Under stress conditions, both platforms---aside from Azure---maintained acceptable error rates below 2\% in every scenario.

When interpreting our findings, it is important to keep in mind that the free-tier services were used. Although this provides helpful insights for small-scale or pilot deployments, a direct comparison of PaaS-to-PaaS or serverless-to-serverless offerings at higher service tiers may yield different results. Future studies should include such comparisons to provide a more thorough perspective on enterprise-scale deployments.

\subsection{Expense Factors}

Cost estimation based on actual API calls and data transferred during our tests indicates that both platforms offer low operating costs for the tested workloads, especially within free-tier limits. Azure's lower per-request pricing structure was the main reason for its 71.9\% lower costs at baseline load. GCP's per-request billing model can be particularly cost-effective for applications with erratic or unpredictable traffic because users only pay for the actual time it takes to process requests. Azure App Service's flat-rate pricing may provide more predictability for workloads that are regular and high in volume.

\subsection{Guidelines for Platform Selection}

The following useful suggestions are supported by our findings:

Choose GCP Cloud Run if you: (1) have experience with containerized deployments and the Google Cloud ecosystem; (2) value fine-grained scaling and per-request billing; or (3) have workloads that are highly variable and susceptible to unpredictable spikes in traffic.

If you need always-on availability without cold starts, have consistent and predictable workloads, or use Microsoft's enterprise tools and services, go with Azure App Service.

A multi-cloud strategy that capitalizes on the advantages of each platform for particular components may be advantageous for businesses with a range of needs.

\subsection{Restrictions}

This study is subject to several limitations. First, the workloads are artificial and may not accurately reflect the nuances of real point-of-sale traffic, despite being representative. Second, results may differ in other locations due to network latency or regional capacity, as all tests were conducted in a single geographic area. Third, our deployments relied on low-tech, straightforward architectures without advanced features like caching or CDN integration that could further optimize cost or performance. Finally, these results are merely a snapshot in time because cloud services are evolving so rapidly.

Despite these limitations, our open-source, code-driven methodology provides researchers and retailers with a transparent and replicable framework for assessing cloud point-of-sale system performance and cost using their own workloads and configurations.

\section{Conclusion}

Using free-tier resources and open-source benchmarking scripts, this paper provides a transparent and repeatable comparison of Google Cloud Platform Cloud Run and Microsoft Azure App Service for point-of-sale (POS) workloads. Researchers and retailers can use this methodology to empirically assess cloud platforms based on throughput, error rates, real-time performance, and estimated operating costs. All results originate from the code outputs.

Our results demonstrate that both GCP and Azure can maintain scalable throughput and low error rates while reliably managing point-of-sale workloads at the tested scales. GCP Cloud Run is particularly well-suited for retailers who deal with highly variable or unpredictable traffic because it showed superior throughput under stress conditions (70.84 TPS vs. 20.14 TPS) and 23.0\% faster response times at baseline load. For retailers with predictable traffic patterns and cost-sensitive operations, Azure App Service, on the other hand, demonstrated 71.9\% lower operational costs and more consistent response times under steady-state loads.

The open-source, code-driven framework created in this study can be used by any retailer to duplicate these evaluations, modify the workload model to fit their setting, and make informed decisions regarding cloud adoption. By using automated measurement and free-tier resources, companies of all sizes can still employ this strategy.

Future studies will assess advanced features like caching, auto-scaling, and CDN integration; they will also look into multi-region deployments for larger retailers; and they will expand this benchmarking framework to include additional cloud providers, such as Amazon Web Services. Future studies will concentrate on automating continuous platform evaluation as services change, incorporating real production workloads in partnership with industry partners, and broadening the methodology to encompass security, compliance, and hybrid or multi-cloud strategies.

Easy-to-use and repeatable benchmarking tools are crucial for maximizing technology investments as the retail industry continues its digital transformation. By creating the framework for continuous, data-driven assessment of cloud platforms, this work assists retailers in striking a balance between cost, performance, and operational resilience in a quickly changing environment.

\bibliographystyle{spmpsci}
\bibliography{references}

\end{document}